\documentclass[preprint]{aastex}
\usepackage{amsmath}
\usepackage{amssymb}
\usepackage{graphicx}
\usepackage{multicol}
\usepackage{multirow}
\usepackage{color}

\begin{document}

\newcommand {\gs} {{\tt{GalSim}}}  
\newcommand {\wf} {WFIRST}  
\newcommand {\wfa} {WFIRST}  
\newcommand {\optical} {{\tt{OpticalPSF}}} 
\newcommand {\gauss} {{\tt{Gaussian}}}

\title{The effect of detector nonlinearity on \wfa\ PSF profiles for weak gravitational lensing measurements}
\author{A. A. Plazas$^{\dagger a}$, C. Shapiro$^{a,b}$, A. Kannawadi$^c$, R. Mandelbaum$^c$, J. Rhodes$^{a,b,d}$, $\&$ R. Smith$^b$  }
\email{$^\dagger$Andres.A.Plazas.Malagon@jpl.nasa.gov}
\affil{$^a$Jet Propulsion Laboratory, California Institute of Technology, 4800 Oak Grove Dr., Pasadena, CA 91109, USA}
\affil{$^b$California Institute of Technology, 1200 E. California Blvd., CA 91125, USA}
\affil{$^c$McWilliams Center for Cosmology, Department of Physics, Carnegie Mellon University, Pittsburgh, PA 15213, USA}
\affil{$^d$Institute for the Physics and Mathematics of the Universe, 5-1-5 Kashiwanoha, Kashiwa, Chiba Prefecture 277-8583, Japan }

\begin{abstract}
Weak gravitational lensing (WL) is one of the most powerful techniques to learn about the dark sector of the universe. To extract the WL signal from astronomical observations, galaxy shapes must be measured and corrected for the point spread function (PSF) of the imaging system with extreme accuracy. Future WL missions\textemdash such as NASA's Wide-Field Infrared Survey Telescope (WFIRST)\textemdash will use a family of hybrid near-infrared CMOS detectors (HAWAII-4RG) that are untested for accurate WL measurements.  Like all image sensors, these devices are subject to conversion gain nonlinearities (voltage response to collected photo-charge) that bias the shape and size of bright objects such as reference stars that are used in PSF determination.
We study this type of detector nonlinearity (NL) and show how to derive requirements on it from \wf\ PSF size and ellipticity requirements. We simulate the PSF optical profiles expected for \wf\, and measure the fractional error in the PSF size ($\Delta R/R$) and the absolute error in the PSF ellipticity ($\Delta e$) as a function of star magnitude and the NL model. For our nominal NL model (a quadratic correction), we find that, uncalibrated, NL can induce an error of $\Delta R/R=1\times10^{-2}$ and $\Delta e_2=1.75\times 10^{-3}$ in the H158 bandpass for the brightest unsaturated stars in \wf. In addition, our simulations show that to limit the bias of $\Delta R/R$ and $\Delta e$ in the H158 band to $\sim 10\%$ of the estimated \wf\ error budget, the quadratic NL model parameter $\beta$ must be calibrated to $\sim1\%$ and $\sim 2.4\%$, respectively. We present a fitting formula that can be used to estimate \wf\ detector NL requirements once a true PSF error budget is established.
\end{abstract}

\section{Introduction}
Weak gravitational lensing (WL) has been identified as a powerful probe of the nature and evolution of the components of the Universe. In particular, \emph{cosmic shear}\textemdash the subtle  distortions of background galaxy shapes by the large-scale structure of the Universe\textemdash constrains the properties of dark matter and dark energy through the measurement of the expansion history and growth of structure of the Universe \citep{refregier03, hoekstra08, kilbinger15}. WL measurements also \textcolor{black}{allow the testing of} the validity of General Relativity that relates the gravitational potential to the matter-energy distribution. Several surveys in the visible part of the spectrum  of $>1000$~deg$^2$ of the sky are currently underway and use the WL signal from hundreds of millions of galaxies as one of their central scientific techniques (\emph{e.g.}, the Dark Energy Survey (DES), \citealt{diehl12}, \citealt{jarvis15}; the Kilo-Degree Survey (KiDS), \citealt{kuijken15}; and the Hyper Suprime-Cam Survey (HSC), \citealt{miyazaki12}). In addition, future ground- and space-based surveys and missions in the visible and near infra-red (NIR) are planned to image more than O($10^9$) galaxies in the next decade (\emph{e.g.}, the Large Synoptic Survey Telescope (LSST), \citealt{ivezic08}; the Euclid spacecraft, \citealt{laureijs11}; and NASA's Wide-Field Infrared Survey Telescope (WFIRST), \citealt{green12}, \citealt{spergel13}, \citealt{spergel15}).

The process of extracting the WL signal from images of the sky, in the presence of intrinsic galaxy ellipticity variations that are $\sim$ 0.4 r.m.s, is highly non-trivial. It must be done through a statistical analysis of large galaxy samples, with a careful control of systematic uncertainties. The dominant signal produced by WL can be described by a local linear transformation of the source image that produces a shear (a complex, spin-2 field of components $\gamma_1$ and $\gamma_2$) and a scalar magnification, both of which have an r.m.s. amplitude of only $\sim 2 \%$ in the case of cosmic shear. Most of the background galaxies usable by WL are at hight redshift (with low signal-to-noise (S/N) ratio) and with a size comparable or smaller than the Point Spread Function (PSF) of the imaging system. Incorrect estimation of the size of the PSF induces a modulation in the signal (multiplicative errors), and errors in the estimation of the PSF ellipticity propagate into asymmetries that produce coherent spurious patterns (additive errors) that mimic the WL signal. Bright stars are commonly used to estimate the PSF, and then this information must be interpolated to the observed galaxy positions to deconvolve the PSF contribution and measure the galaxy shape (in the form of a complex ellipticity $e=e_1 + i e_2$) to estimate the shear field.\footnote{Most of the shape measurement algorithms to date rely on the accurate measurements of galaxy shapes to produce an estimator of the WL shear field $(\gamma_1, \gamma_2)$. However, recent algorithms propose skipping this step and creating a direct shear estimator through Bayesian analysis (\citealt{miller13}, \citealt{bernstein14}, \citealt{bernstein15}, \citealt{schneider15}, \citealt{alsing16}).} This interpolation step introduces systematic errors if the information inferred from the stars does not fully constrain the PSF at the galaxy position with the required accuracy. In order not to bias the determination of dark energy and other cosmological parameters in Stage IV surveys (in the language of \citealt{albrecht06}), the ellipticity and relative size of the PSF must be known to an accuracy of O(10$^{-3}$) (\citealt{huterer06}, \citealt{amara08}, \citealt{paulin08}, \citealt{paulin09}, \citealt{massey13}, \citealt{cropper13}) or better ($4.7 \times 10^{-4}$ for the knowledge of the \wfa\ PSF ellipticity, \citealt{spergel13}).

Systematic errors that originate from a telescope's detectors (image sensors) introduce biases in astronomical observables such as photometry and astrometry that propagate into shear measurement biases. These type of errors have been extensively studied in the case of thick, fully-depleted, high-resistivity Charge Coupled-Devices (CCDs), which are the detectors of choice for many current and planned surveys such as DES, HSC, and LSST (\emph{e.g.}, \citealt{stubbs14}, \citealt{plazas14}, \citealt{gruen15}), as well as in other types of CCDs (\emph{e.g.}, \citealt{prodhomme14}, \citealt{niemi15}). It is of great importance to quantify the impact of these sensor effects on the inference of cosmological parameters, in particular through WL (\citealt{jarvis14}, \citealt{mandelbaum14b}, \citealt{meyers14}). Future missions such as the James Webb Space Telescope and \wfa\ will utilize a family of near-infrared detectors that are also subject to effects such as nonlinearity, reciprocity failure (\citealt{bohlin05}, \citealt{biesiadzinski11}), interpixel capacitance (IPC; \citealt{mccullough08}, \citealt{kannawadi15}), and persistence (\citealt{smith08}). These effects can imprint biases on weak lensing shape measurements if not taken into account. 

In this paper we study the effect of nonlinear detector conversion gain (voltage response to collected photo-charge) on PSF size and ellipticity in the context of the NIR detectors that will be used by NASA's \wf\ mission. This type of detector nonlinearity (NL) will tend to attenuate the measured flux in bright stars, broadening the inferred PSF. In other words, NL preferentially depresses the flux in the core of the PSF relative to the wings, thus complicating its deconvolution from the observed galaxy image, which itself is fainter and less subject to the effects of NL. In addition, even though NL does not induce a spurious ellipticity by itself, it modifies the PSF ellipticity if the PSF is anisotropic. Our analysis is also useful to set preliminary requirements on NL for these sensors. Once characterized, NL can be corrected in each image, and remaining residuals will depend on the accuracy in the knowledge of the NL parameters and their spatial variation. We use the {\tt{python}}/{\tt{C++}} code \gs\footnote{\url{https://github.com/GalSim-developers/GalSim}, \url{https://wfirst.ipac.caltech.edu/sims/Code.html}} \citep{rowe15} to simulate \wfa\ PSF profiles and to analyze the impact of NL on PSF size and ellipticity. 

In Section 2 we summarize the main characteristics of the NIR detectors that will be used in \wfa\, and describe NL. In Section 3 we describe the simulations we create to study NL for \wfa\ PSF profiles. Section 4 presents our main results on fractional errors in size and absolute errors in ellipticity caused by NL, as function of relevant parameters such as the model parameters and PSF magnitude. We also study the effect of the spatial variability of the NL model across the pixel array. We conclude in Section 5 with a discussion of our results and how they can be used in the derivation of NIR detector specifications to satisfy WL accuracy requirements.  

\section{Voltage nonlinearity in the NIR detectors of \wfa}
\label{section:NL}

The \wfa\ mission will use a 2.4 m telescope equipped with a Wide Field Instrument (WFI) with 6 bandpass filters: Z087, Y106, J129, W149, H158, and F184 (\citealt{spergel15}).\footnote{In addition to a integral field unit and a coronagraph for supernovae and exoplanet studies, respectively.} The WFI will perform a high-latitude survey (HLS), imaging over an area of 2200 deg$^2$ in four NIR ($\sim0.92$\textendash $2.00\ \mu$m) bands (Y106, J129, H158, and F184) down to a 5$\sigma$ point-source AB magnitude of 26.7 in the J129 band. The weak lensing program in the HLS will measure shapes of about 380 million galaxies in the J129, H158, and F184 bands (\citealt{spergel15}). 

The WFI possesses a wide-field channel that has a Focal Plane Assembly (FPA) of 18 $4 k\times 4k$ HgCdTe (mercury, cadmium, and telluride) NIR detectors, arranged in a $6\times3$ layout and with a pixel size and scale of 10 $\mu$m and 0.11 arcseconds per pixel, respectively. The HgCdTe NIR detectors are manufactured by Teledyne Imaging Systems, and are part of a family of detectors known as Hawaii-XRG (HXRG), where X denotes the detector width in thousands of pixels\footnote{HAWAII stands for HgCdTe Astronomical Wide Area Infrared Imager, and RG stands for ``Reference pixels and Guide mode''.}. 

The detector arrays are fabricated with a hybrid complementary metal-oxide-semiconductor (CMOS) architecture, which combines the qualities of HgCdTe to detect infra-red light (\emph{e.g.}, altering the relative molar contributions of mercury and cadmium allows one to tune the band gap by up to an order of magnitude) and the advanced readout performance of integrated circuits. Light is absorbed, converted to charge through the photoelectric effect, and collected by electric fields generated by a reverse-biased p-n junction in the detector layer.  The charge per pixel is then converted to a voltage and amplified through a source follower. This operation is performed in the silicon readout integrated circuit (ROIC) layer, which is connected to the HgCdTe detection layer by indium interconnects (one indium bump per pixel). Finally, the ROIC transfers the signal (and for this it is also known as ``multiplexer") to the off-chip electronics at the edge of the FPA, where it is digitized through analog-to-digital converters \citep{beletic08}.

An ideal detector would produce a measurable signal that is proportional to the detected photons. However, there are several places in the signal chain where this expected linearity is not realized, and the conversion of charge to measured voltage (or digital numbers) becomes non-linear. Each pixel's p-n junction acts as a parallel-plate capacitor, and as charge accumulates the depletion region narrows, causing a deviation from linearity of the charge-to-voltage conversion relation. Nonlinearity can also be introduced through the electronic gain of the ROIC. Furthermore, the charge accumulation rate might be a function of the photon-accumulation rate, an effect known as count-rate nonlinearity or reciprocity failure (RF) (\citealt{smith08}, \citealt{biesiadzinski11}). 
The first two types of nonlinearity depend only on fluence (integrated signal) as opposed to RF, which is flux dependent.  They can be analyzed together in a single transfer function typically called ``nonlinearity" (NL). We study the impacts of NL (more relevant at high signals) on PSF measurements in this paper, while we leave investigations on the consequences of RF (relevant at lower signals) on WL measurements for future work. 

Assuming that the dominant contribution to nonlinearity is the varying capacitance of the pixel p-n junction, we find that the correction to the detector signal is well-approximated by a quadratic term:
\begin{align}
S(Q)= Q - \beta Q ^2
\label{NL}
\end{align}
Here, $Q$ is the true number of elementary charges collected in the pixel, $S$ is the number inferred from the voltage change at the sense node, and $\beta$ is a constant.  To estimate $\beta$, we compute $Q(V)=C(V)*V/q_e$, where $q_e$ is the elementary charge, and the total capacitance is given by the varying junction capacitance plus a constant $C(V) = C_{\rm jn}(V) + C_{\rm fix}$. The junction capacitance varies as (\citealt{mccaughrean87})
\begin{equation}
C_{\rm jn}(V) \propto (1+V/V_{bi})^{-1/2} \;.
\end{equation}
Here, $V_{bi}$ is the ``built in'' potential of the junction and $V=V_{\rm DSUB}-V_{\rm RESET}+\delta V$, where $V_{\rm DSUB}$ is the constant potential at the diode cathode, $V_{\rm RESET}$ is the initial potential at the anode, and $\delta V$ is the change in anode potential due to accumulated photocharge.  As an example, we substitute measurements of a 2.4$\mu$m cutoff H2RG by Finger 2006:\footnote{\url{https://www.eso.org/sci/meetings/2006/neon-2006/Finger.pdf}}
\begin{eqnarray*}
V_{\rm bi} &=& 0.412\ {\mbox V} \\
V_{\rm DSUB} &=& 1\ {\mbox V} \\
V_{\rm RESET} &=& 0.5\ {\mbox V} \\
C_{\rm fix} &=& 17.8\ {\mbox fF} \\
C_{\rm jn}(V=.912\ {\mbox V}) &=& 30\ {\mbox fF}
\end{eqnarray*}
Using these parameters to compute $Q(V)$ for $ 0\le \delta V \le 0.3$ V (corresponding to maximum $Q$=60115) and then inverting to find $S(Q)$, we find it is well-fit (to 0.1\% or better) by Eq. \ref{NL} with $\beta=1.18 \times 10^{-6}$. In practice, HXRG calibrations have included additional polynomial parameters which can reduce residuals, extend the range of valid $Q$, and account for additional nonlinear effects (\citealt{hilbert04}, \citealt{hilbert08}, \citealt{hilbert14}); however, the additional parameters are highly degenerate, resulting in large variances in the fitted values.  For our purposes, it suffices to analyze the shape-distorting effects of nonlinearity using a single parameter, $\beta$, which encapsulates most of the effect. In this paper we chose a nominal value of $\beta_0=5\times 10^{-7}$, which is near the midpoint of the measured range for $\beta$ in \citealt{hilbert14}. 

To illustrate the effect of NL on photometry, consider nearly saturated stars in the HLS.  Using PSF profiles with a simulated AB magnitude up to 18.3 for an exposure time of 168.1 seconds (expected for the HLS), the total flux in each band is shown in Table 1, along with the peak pixel value. When the profile is drawn on a pixelated postage stamp with a scale of $p=0.11$ arcseconds and placed on the center of the pixel,\footnote{Note that the peak value is a function of the PSF profile centroid location within the pixel, as well as pixel resolution. When drawing the PSF profiles in our simulations, we randomize the PSF centroid within the native scale pixel, as described in Section 3.} this particular magnitude produces a peak charge of $\sim 1\times 10^5$ e$^{-}$ (Y106 band, see Table 1), which represents about 90\% of the typical pixel full well value of $1.1 \times 10^5$ e$^{-}$.\footnote{Dave Content, private communication.}
At this level of charge, the signal attenuation due to NL for the nominal $\beta_0$ value is about 5\% (Eq. \ref{NL}). 

Calculations with the {\tt{Trilegal}} galaxy model\footnote{\url{http://stev.oapd.inaf.it/cgi-bin/trilegal}} \citep{girardi12} show that there will be approximately 20 stars at or brighter than this magnitude (18.3) and flux level per detector.\footnote{We thank Christopher Hirata, who wrote the code to perform these calculations. The NIR filter sets used in {\tt{Trilegal}} had to be converted from the vega to the AB magnitude system. The code was run at the South Galactic Pole with the SDSS and 2MASS ($ugriz$ and $JHK_\mathrm{s}$) filters in 1 deg$^2$ and interpolated to WFIRST filter centers, which ignores the fact that stars have spectral structure in the NIR, but gives a result good to about $10\%$. The number of stars quoted in the text is expected to be higher at moderate galactic latitudes.} We assume that galaxies will have about two orders of magnitude fewer total electrons than bright stars, and therefore for this quadratic model, star shapes are distorted by NL and galaxies are (approximately) not, which would result in an incorrect PSF correction if not accurately calibrated. Thus our goal is apply NL to simulated \wfa\ PSF profiles and quantify the impact on PSF properties such as size and shape. 

\section{Methods}
\label{methods}
\subsection {Simulations}
We use the publicly available {\tt{GalSim}} code (v1.3) to simulate the impact of NL on the \wfa\ PSF shape and size. {\tt{GalSim}} is a {\tt{python/C++}} open-source code that allows the user to create simulations of astronomical objects, and it was developed by the weak lensing community to investigate shape measurement algorithms and systematics. 

\citet{kannawadi15} have developed within \gs\ v1.3 a \wfa\ module called ``{\tt{galsim.\\ wfirst}}", which allows the simulation of a PSF profile\footnote{By calling the {\tt{galsim.wfirst.getPSF}} routine.} according to the optical design characteristics of the \wfa\ WFI \citep{pasquale14}\footnote{\citealt{pasquale14} discuss the so-called ``Cycle 4" optical design, whereas the \gs\ \wfa\ module\textemdash used in this work\textemdash uses files corresponding to ``Cycle 5".} 

For this work, we have created simulations in the four bands of the HLS.  A central circular obscuration ($30 \%$; linear in value) and six support struts are included as well. 

The module generates PSF models that do not include pointing jitter nor charge diffusion. These effects could be added to the profile (\emph{e.g.,} by means of an extra convolution with a Gaussian profile in the case of diffusion), but we did not include them in our simulations. Their effect would be to create a slightly larger PSF, reducing the impact of NL. Their omission makes our results slightly conservative.  

The ellipticity of the WFI PSF varies over the field of view due to optical aberrations. To make our results conservative, we simulate only detector \#18, whose PSF was determined to have the largest ellipticity (see Section 3.2 for a description of the shape measurement method used) across all bands among all the detectors (Table 1). We also evaluate the PSF at the mean wavelength weighted by each bandpass or effective wavelength.\footnote{This is called an achromatic {\tt{galsim.OpticalPSF}} object in \gs.}
We then form an effective PSF by convolving the profile with a 2D top-hat profile of length equal to the nominal angular scale of the \wfa\ WFI ($0.11$ arcseconds per pixel). 

\begin{table}[!htb]
\centering
\begin{tabular}{ |c| c | c| c| c| c| c | c| c|}
\hline
\multirow{2}{*}{Band} & Min. $\lambda$ & Max. $\lambda$ & $\lambda_{\text{eff}}$ & $b(\lambda)$ & peak value  & e$_1$ & e$_2$ & $\sigma$ (pix) \\
& ($\mu$m) & ($\mu$m) & ($\mu$m) & ($\times10^5$) (e$^{-}$) & ($\times 10^{5}$) (e$^{-}$) & (\#18) & (\#18) & (\#18)  \\
\hline 
Y106 & 0.900 & 1.230 & 1.061 & 2.7621 &  1.00237  &-0.0163  & 0.2035 & 1.7020 \\
J129  & 1.095 & 1.500 & 1.292 & 2.8267 &  0.89742 & -0.0127  & 0.1325 & 1.717 \\
H158 & 1.340 & 1.830 & 1.577 & 2.7922 &  0.38654 & -0.0089 & 0.0802 & 1.832 \\
F184 & 1.630 & 2.060 & 1.837 & 1.8346 &  0.71890 & -0.0071 & 0.0550 & 1.995 \\
\hline
\end{tabular}
\caption{The first column lists the four bands that will be used in the HLS. For weak lensing analysis, multi-band shape measurement will be done in bands J129, H158, and F184. Columns 2, 3, and 4 show their minimum, maximum, and effective wavelengths, respectively (from \citealt{kannawadi15}). Column 5 shows the baseline total flux (in electrons) $b(\lambda)$ of Eq. \ref{flux} at AB magnitude 18.3 (at 168.1 seconds of exposure time) in each band as calculated in \gs, while column 6 shows the peak value for each profile when rendered on a postage stamp at the native pixel scale ($p=0.11$ arcseconds per pixel) at that same magnitude and with the PSF centroid at the center of the pixel. The last three columns show the ellipticity components and size of the \wfa\ PSF profile (drawn a a resolution of $p/N$, with $N=3$) in detector number 18 as calculated by the adaptive moments routine in \gs. Chip \#18 was found to possess the largest values of absolute PSF ellipticity.}
\label{table1}
\end{table}

The WFI PSF is undersampled by design. In order to maximize the field of view, detectors in instruments of space missions are usually built with a physical size that results in undersampled images, which fail to satisfy the Nyquist-Shannon criterium for the maximum band limit set by the optical response of the system, and therefore produce aliased images.\footnote{The Nyquist-Shannon criterium states that the sampling interval $p$ must satisfy $p < 1/(2 u_{max})$, where $u_{max}$ is the highest frequency in the signal, in order to avoid aliasing.} In general, it is not possible to recover all the information of a continuous function from a discrete sample of points if the image is aliased, and measurements of astronomical object's properties such as magnitude and shape will be erroneous (\citealt{lauer99a}, \citealt{fruchter11}, \citealt{rhodes07}).  Undersampled data can be fit to a model at the risk of introducing model bias.  In practice, we find that our shape measurement algorithm (see Section 3.2 below) performs poorly on undersampled images\textemdash when most light is concentrated in one or a few pixels, the image ellipticity is poorly constrained.  Thus we must simulate the effect of NL on oversampled images.

To overcome the problem of undersampling in real data, multiple dithered exposures are taken, and then processed by an image combination algorithm (\citealt{lauer99b}, \citealt{fruchter02}, \citealt{bertin06}, \citealt{rowe11}) in order to produce an oversampled image that satisfies the Nyquist-Shannon criterium. To study NL with repeated simulations, however, such external software is computationally expensive.  Instead, we approximate oversampled data by rendering the convolved PSF profile\textemdash including the pixel response at its native scale\textemdash as a \gs\ object at a high resolution,\footnote{This is done by calling the {\tt{galsim.GSObject.drawImage}} method with option``{\tt{method=no\char`_pixel}}".} setting the parameter {\tt{scale}} to $p/N$, where $N$ is a positive integer and $p$ is the native pixel scale. 
 Defining $Q\equiv1/p\times u_{\text{max}}$ as the sampling factor ($Q<2$, $Q=2$, and $Q\geq2$ represent under-, critically, and over-sampled images, respectively), we see that in order to produce an over-sampled image ($Q\geq2$), the Nyquist-Shannon criterium implies that $N$ must be given by (see \emph{e.g.}, \citealt{marks09}, \citealt{shapiro13})

\begin{align}
N=\frac{2p}{\lambda_{min} F}
\label{nimages}
\end{align}
where $p$ is the pixel size, $\lambda_{min}$ is the shortest wavelength in a given filter, and $F$ is the the f-number of the telescope. In the case of the \wfa\ telescope, $p=10\ \mu$m and $F=7.8$, resulting in $N\geq3$ ($N\geq2$) for the J129 (H158) band. Thus, we have chosen to set $N=3$ in our simulations to ensure oversampling.

Note that the NL effect on the reconstructed image from dithering real exposures will depend on the centroid of the source in the dithered inputs: an exposure where the PSF peaks at a pixel center will have higher NL than an exposure where the peak is spread more evenly over 2-4 pixels. To mimic this behavior in our approximation, we create a PSF profile whose centroid coordinates are random numbers uniformly distributed over the size of a native detector pixel. We then render the profile at a higher resolution $p/N$ and measure its average size and shape over 100 realizations. With $N$=3, the variation of the NL correction in the pixels of the oversampled image should be similar to the variations in the 5-8 exposures combined using the WFIRST dither strategy.  In practice, we find that the scatters of the shape measurements over these realizations are negligible relative to the means; therefore, we expect that the effect of NL on the shape of an oversampled image is insensitive to the precise dither pattern.

Notice that this method to approximate oversampled images can only be used in the case of sensor effects such as NL, which depend on each pixel individually. In the case of other effects that correlate the signal in one pixel with the signal of its neighbors (such as IPC), the effect should be applied to the native scale of the detector and then a properly sampled image should be generated through adequate combination of several dithered images before performing any analysis (see for instance the ``interleaving" algorithm of \citealt{kannawadi15}, also included in \gs). 

\begin{figure}[!ht]
\centering
\resizebox{\hsize}{!}{\includegraphics{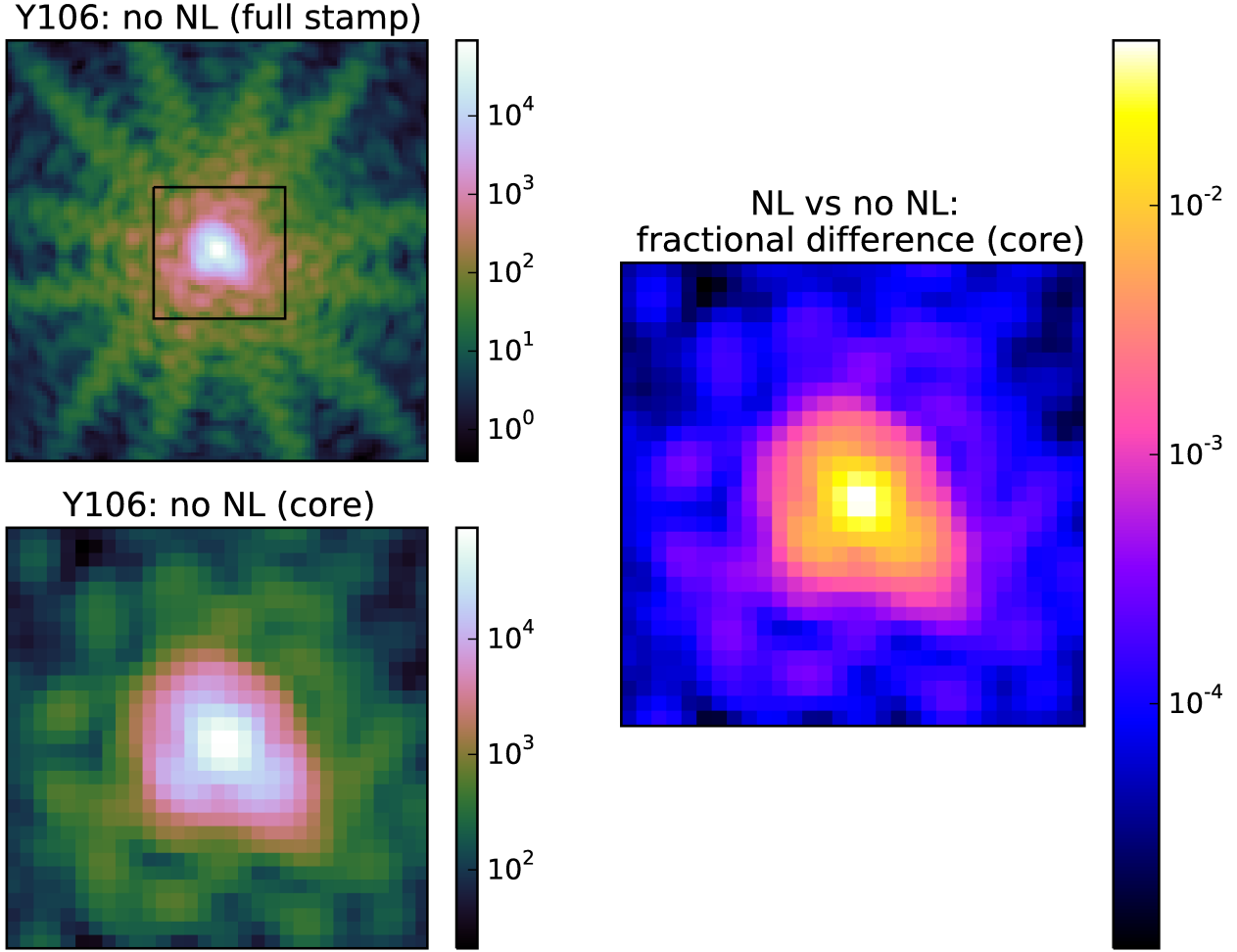}}
\caption{Example of the \wfa\ PSF profiles in the Y106 band created by the {\tt{galsim.wfirst.getPSF}} method. The profile is drawn on a postage stamp with a size of $1.5k \times1.5k$ pixels ($k=64$) and a resolution of $p/N$, with $p=0.11$ arcseconds and $N=3$. Before drawing the profiles, they are first convolved with a pixel response of size $p$ and given a flux obtained through the use of the {\tt{galsim.wfirst.getBandpasses (AB\char`_zeropoint=True)}} routine at an AB magnitude of 18.3 (at an exposure time of 168.1 seconds). However, to preserve the correct response to NL, the higher-resolution image has $N^2$ more total flux. The NL effect is applied at the high-resolution pixel scale $p/N$, but the centroid of the profile is randomized within the native pixel scale $p$ (in this example the centroid and the pixel center coincide). The upper left panel shows the full postage stamp PSF image without the NL applied, while the lower left panel shows a zoom into the core (squared central region in upper left panel) of 30 by 30 high-resolution pixels. The right-hand side image shows the fractional difference between the PSF without the NL applied and a PSF with NL using $\beta_0$ for the model parameter.} 
\label{f1}
\end{figure}

The PSF profiles are drawn into squared postage stamps of size $1.5k$ pixels ($k=64$), large enough to ensure that most ($\sim 96\%$) of the total PSF flux is contained in the stamp. Each profile object was assigned a total flux of 

\begin{align}
f=b(\lambda)\times10^{0.4 (18.3-m)} \ \mathrm{e^-}, 
\label{flux}
\end{align}
Eq. \ref{flux} gives the number of source counts per exposure, and $m$ represents the AB magnitude of the object for a given band. $b(\lambda)$ e$^{-}$ is a baseline flux at AB magnitude $m=18.3$ for a exposure time of $168.1$.\footnote{The magnitudes are determined by using the routine {\tt{galsim.wfirst.getBandpasses (AB\char`_zeropoint=True).}}}

We have neglected the main sources of noise that would affect the HLS\textemdash zodiacal background, thermal emission, and read noise\textemdash which would make all images slightly more nonlinear. Calculations performed with the \wfa\ Exposure Time Calculator v.14\footnote{\url{https://wfirst.ipac.caltech.edu/sims/tools/wfDepc/wfDepc.html}} \citep{hirata12} in weak lensing mode (ETC-WL) by \citealt{spergel15} in the F184 band (a conservative case) show that the combined contribution due to these backgrounds sources would be approximately 130 $e^-$ per pixel for a $174$ seconds exposure (comparable to the $168.1$ seconds of exposure time assumed in this work), which creates negligible corrections in our simulations for a NL parameter $\beta$ of order $10^{-6}$.

Total and peak flux values at each filter per postage stamp are shown in Table 1. The peaks are upper limits estimated by assuming the PSF centroid is centered on a pixel.  
The brightest magnitude used was AB $m=18.3$, based on the peak value of the PSF profile in the Y106 band when rendered into a postage stamp at the native pixel scale and placed at the center of the pixel: $\sim 1.00237\times10^{5}$ electrons, representing about $90\%$ of the typical full well depth in the pixels of the H4RG detectors\textemdash$\ \sim1.1\times10^{5}$ electrons.
Recall that for each postage stamp, we select a centroid with coordinates randomly chosen from a uniform distributed across the native scale pixel. In addition, since NL is a function of signal, and since \gs\ conserves total flux when changing pixel scales, the total flux in the higher resolution image must be multiplied by a factor of $N^2$ to preserve the appropriate response per pixel to this effect, correcting for the fact that the new image has a factor of $N^2$ more pixels than the one created at the native scale. 

For simplicity of analysis, the postage stamps are noiseless and no other sensor effect is applied. After convolving the PSF profile with a pixel of the size of the native scale and rendering the image at the high-resolution scale, NL is applied by using Eq. \ref{NL}.\footnote{{\tt{galsim.image.applyNonlinearity.}}}

Fig. \ref{f1} shows an example of the PSF profiles and postage stamps created for our simulations (in the Y106 band). The effect of NL (at the nominal $\beta_0$) is small, and the difference image reveals that the attenuation in the flux is of the order of a few percent, mainly for the larger signals found at the core of the PSF.  \citealt{kannawadi15} present more details on the {\tt{galsim.wfirst}} module, along with examples of the PSF profiles that can be generated in all 6 \wfa\ filters. 
\subsection {PSF size and shape measurement}
The accurate determination of PSF properties such as size and ellipticity is crucial to avoid the propagation of systematic biases in cosmological parameters through the use of weak gravitational lensing (\citealt{paulin08}, \citealt{paulin09}, \citealt{massey13}, \citealt{cropper13}). In general, the problem of galaxy and PSF shape measurements for accurate weak lensing is non-trivial, and even when the PSF is perfectly known, shape measurement algorithms can introduce biases. Several shape measurements algorithms\textemdash ranging from model-fitting methods to particular combinations of weighted central moments and bayesian techniques\textemdash have been and are being investigated in order to produce accurate shear estimators that satisfy the requirements of current and future WL surveys \citep{mandelbaum15}.

To measure the profile shapes and size, we use the adaptive moments method (\citealt{bernstein02}, \citealt{hirata03}).\footnote{Already implemented in \gs\ as {\tt{galsim.hsm.FindAdaptiveMom().}} }. Adaptive moments are effectively weighted by an elliptical Gaussian. At first they are calculated by computing moments weighted by a circular Gaussian with some arbitrary size. Then the output moments are used to define a new elliptical Gaussian that will act as a new weight function. The process is iteratively repeated until the output moments are the same as those of the weight function. The ellipticity $e=e_1 + ie_2$ and size $R$ are then defined as

\begin{align}
e_1&=\frac{M_{xx} - M_{yy}}{M_{xx}+M_{yy}} \\
e_2&=\frac{2 M_{xy}}{M_{xx}+M_{yy}} \\
R&=\det[\mathbf{M}]^{1/4}
 \end{align}
where the centroid $\mathbf{\bar{x}}$ and moment matrix $\mathbf{M}$ of an image are defined as
\begin{align}
\mathbf{\bar{x}}&=\frac{\int d^2\mathbf{x}\ w(\mathbf{x}) \mathbf{x} I(\mathbf{x})}{\int d^2 \mathbf{x}\ w(\mathbf{x})I(\mathbf{x})} \\
M_{ij}&=\frac{\int d^2\mathbf{x}\ (\mathbf{x} - \mathbf{\bar{x}})_i  (\mathbf{x} - \mathbf{\bar{x}})_j w(\mathbf{x}) I(\mathbf{x})} {\int d^2\mathbf{x}\ w(\mathbf{x})I(\mathbf{x})}
\end{align}
for an elliptical Gaussian weight function $w(\mathbf{x})$. 
We note that adaptive moments are particularly sensitive to the core of the PSF when the PSF profile is diffraction-limited, and therefore they should be particularly sensitive to NL effects.

\subsection{Changes in size and ellipticity induced by NL}

We quantify the effect of nonlinearity by measuring the fractional change in size and the absolute change in ellipticity of the PSF profiles,\footnote{In WL, before measuring a given galaxy shape the PSF has to be corrected. By propagating the errors in the determination of the size and ellipticity of the PSF, it can be shown (see, \emph{e.g.}, \citealt{paulin08}) that the error in the measurement of the source galaxy ellipticity is\textemdash to first order\textemdash given by a linear combination of the fractional PSF size error and the absolute PSF ellipticity error, which are the metrics we have chosen to study in this work.} in the 4 filters of the HLS of \wfa\, and for several values of the NL model parameter $\beta$. We calculate the quantities $\Delta e_1$, $\Delta e_2$, and $\frac{\Delta R}{R}$, which are defined as the difference between the measured ellipticity or size after the effect (NL) is applied and the reference values measured before (represented by the subscript ``$0$") the application of NL:

\begin{align}
\Delta e_{i} &\equiv e_{i} - e_{i,0},\ \ \ i\in[1,2]  
\label{delta_e}
\end{align}
\begin{align}
\frac{\Delta R}{R} &\equiv \frac {{R} - R_{0}}  { R_{0}}
\label{delta_r}
\end{align}
In this way, we are less sensitive to the details (and possible biases) of the shape measurement algorithm, since we only care about relative changes induced by the detector effect. 

The basic simulation process is summarized by the following steps:  
\begin{itemize}
\item[1.] Create a \wfa\ PSF surface brightness profile with a given flux as prescribed by Eq. \ref{flux}, and convolve the PSF profile with a top-hat pixel response with the size of the native scale of the \wfa\ FPA ($p=0.11$ arcseconds per pixel) to produce an effective PSF. 
\item[2.] Sample the effective PSF profile on a 2D grid to draw a noiseless postage stamp of size $1.5k$ by $1.5k$ ($k=64$) pixels at a higher resolution of of $p/N$, with $N=3$, and multiply the resulting image by $N^2$. In this step, NL is not yet applied, but it will be done so in a later step, so the flux still needs to be adjusted. The centroid of the profile is randomly selected from a uniform 2D distribution within the size of a native pixel.
\item[3.] Create another image of the effective PSF at a higher resolution and with the flux adjusted as in step 2. 
Apply the nonlinearity to the postage stamps according to the transformation $I \mapsto I - \beta I^2$ (\emph{c.f.} Eq. \ref{NL}).
\item[4.] Use the adaptive moments algorithm to measure the shape $e_0=(e_{1,0}, e_{2,0})$ and size $R_0$ of the profile without NL to have as baseline reference.
\item[5.] Measure the shape and size of the object with the sensor effect applied, and calculate the quantities $\Delta e_1$, $\Delta e_2$, and $\frac{\Delta R}{R}$, as defined in Eqs. \ref{delta_e} and \ref{delta_r}. 
\item [6.] Repeat steps 1 to 5, averaging over 100 centroid realizations. The size and ellipticity values reported will be the mean and the standard deviation over these realizations (see Fig. \ref{f2} and Fig. \ref{f3}).
\item[7.] To study the impact of spatial variability in $\beta$, repeat steps 1 to 6 and assume that the model parameter $\beta$ for each high-resolution pixel can be drawn from a Gaussian distribution with a particular mean $\beta$ and variance $\sigma_{\beta}^2$. Over $M$ realizations (for a fixed centroid), calculate the dispersion values $\sigma_{\Delta e_1}$, $\sigma_{\Delta e_2}$, and $\sigma_{\Delta R/R}$ as a function of $\sigma_{\beta}$. In our simulations we use $M=100$ (see Fig. \ref{f4} and Fig. \ref{f5}). 

\end{itemize}
\section{Results}
\label{results}
\subsection{Biases in ellipticity and size}

\begin{figure}[!h]
\centering
\resizebox{\hsize}{!}{\includegraphics{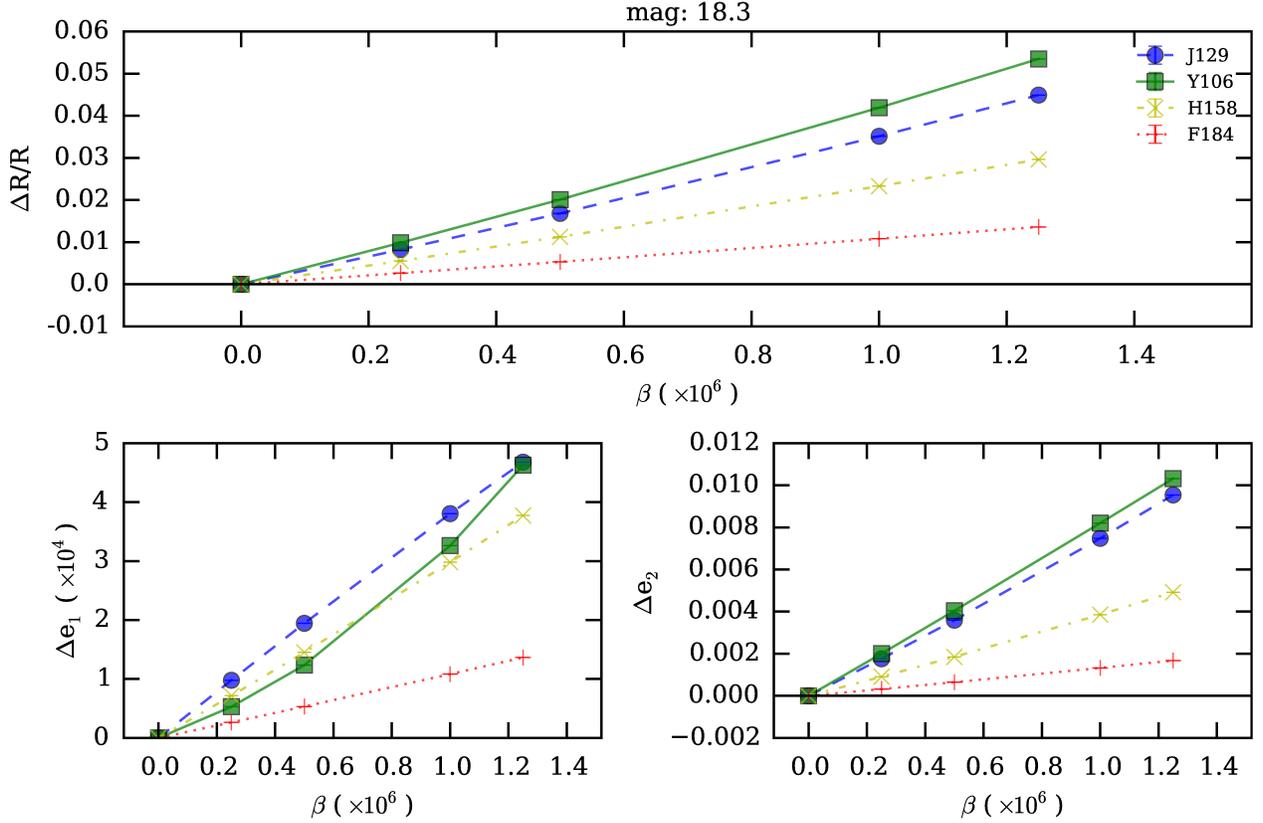}}
\caption{Fractional error in PSF size (upper panel) and absolute error in PSF ellipticity components (lower panels) as a function of the mean nonlinearity parameter $\beta$ for the \wfa\ PSF, in four the four HSL filters (J129, Y106, H158, and F184), and an AB magnitude of 18.3 at 168.1 seconds of exposure time. 
Each point is the mean over 100 realizations of uniformly distributed random centroid shifts within the high resolution images ($N=3$).
The standard deviations of the realizations are negligible.}
\label{f2}
\end{figure}

Fig. \ref{f2} shows the fractional change in PSF size and the absolute error in PSF ellipticity as a function of the mean nonlinearity parameter $\beta$, for different bandpass filters at fixed AB magnitude of 18.3 in each band, consistent with the magnitude of nearly saturated stars in the HLS. We note that the flux in each band will depend on the particular stellar spectral energy distribution (SED), and we would need to specify a particular SED to go from one reference band to other bands. For simplicity, we are performing our calculations based on a grid of AB magnitudes in different filters. 

\begin{figure}[!h]
\centering
\resizebox{\hsize}{!}{\includegraphics{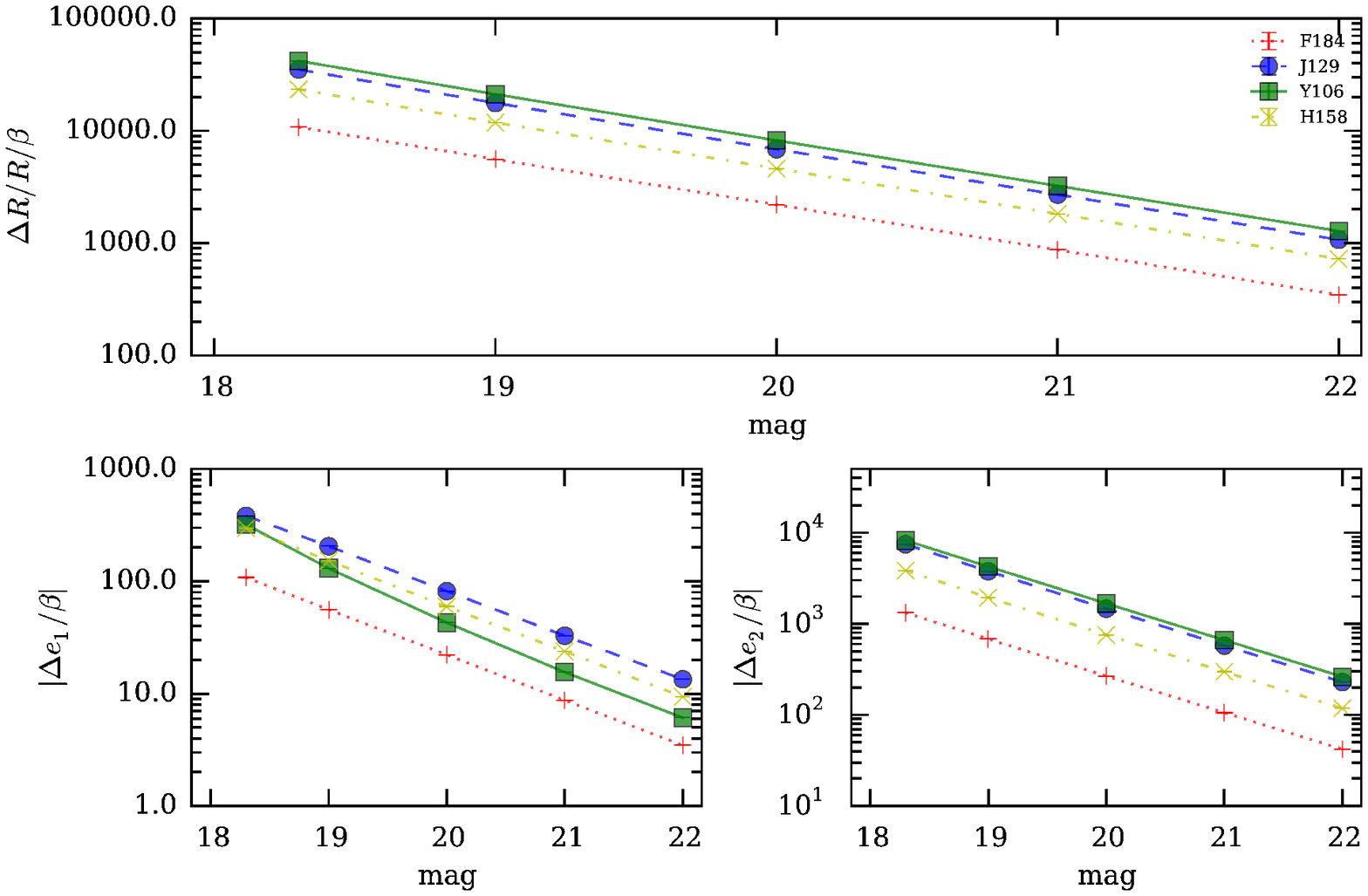}}
\caption{Fractional error in PSF size and absolute error in PSF ellipticity components normalized by the NL model parameter $\beta$, as a function of magnitude for the four HSL filters (J129, Y106, H158, and F184). The ordinate axis in each plot represents the slope of the linear relationships in Fig. \ref{f2}, derived by linear fitting of points around the vicinity of $\beta_0$ in each curve of that figure.}
\label{f3}
\end{figure}

In the H158 band, the mean nominal value of $\beta_0=5\times10^{-7}/\mathrm{e^-}$ induces errors in the size and ellipticity of about $1\times10^{-2}$ and $2\times10^{-3}$ respectively, larger than the required values of $10^{-3}$ and $4.5\times10^{-4}$ on the knowledge of the size and ellipticity of the PSF in order not to bias cosmological parameter inferences from WL experiments (Section1). As the $\beta$ parameter is increased, the amplitude of the errors increases approximately in a linear manner within the domain of $\beta$ values considered. 
 
Since $\Delta R/R$ and $\Delta e$ are approximately linear in $\beta$, we can condense this information by simply plotting the slope for various filters and star magnitudes. This is shown in Fig. \ref{f3}, which presents $\Delta R/R/\beta$ and $\Delta e/\beta$ (the slopes in Fig. \ref{f2}) vs $m$ for each of the four filters of the HLS. From Fig. \ref{f3}, it is possible to estimate the precision to which $\beta$ would have to be calibrated in order to limit the relative size and ellipticity bias of a star with a given magnitude. In particular, letting $(\beta - \beta_0)/\beta_0 \equiv \Delta \beta / \beta_0$ represent the fractional error in the measurement of a given value of $\beta_0$, we have: 

\begin{align}
\frac{\Delta R/R}{\Delta \beta} = c \ \ \ \  \Rightarrow \ \ \ \ 
\frac{\Delta \beta}{\beta_0} = \frac{\Delta R/R}{c \beta_0}
\label{calib}
\end{align}
In Eq. \ref{calib}, $c$ represents the ordinate value in Fig. \ref{f3} for a given magnitude, and we have replaced $\Delta R/R/\beta$ by $\Delta R/R/ \Delta \beta$ since the linearity in the bias in $\Delta R/R$ and $\Delta e$ vs $\beta$ for all magnitudes makes the choice of the expansion point of the approximation by Taylor expansion unimportant (\emph{i.e.}, it does not matter if the expansion is around $\beta=0$ or $\beta=\beta_0$). An analogous equation to Eq. \ref{calib} can be written for the error in the ellipticity if $\Delta R/R$ is replaced by $\Delta e$. 

Under these conditions, Fig. \ref{f3} shows that for the brightest stars in our range, to limit the bias of $\Delta R/R$ ($\Delta e$) in the H158 band to $10^{-4}$  ($4.7\times 10^{-5}$)\textemdash about $10\%$ of the estimated \wf\ error budgets, $\beta$ must be calibrated to $\sim1\%$ ($\sim 2.4\%$) (using $c\sim2\times10^4$ at $m=18.3$ from the upper panel of Fig. \ref{f3}, $c\sim4000$ at $m=18.3$ from the lower right panel of the same figure, and $\beta_0=5\times10^{-7}$). 
Alternatively, this calculation can be used to convert between NL calibration requirements, shape measurement error requirements, and the minimum useable star magnitude: given any two of these, we can obtain the third.  E.g., if NL calibration precision reaches some practical limit, then given a tolerance on shape measurement error, one can find the star magnitude for which the shape measurement bias due to NL matches the tolerance (brighter stars cannot be used to measure the PSF).  Our choice of error tolerance here is only for illustration, and Eq. \ref{calib} and Fig. \ref{f3} should be used to derive detector requirements once true PSF requirements are known.

The trends in Fig. \ref{f3} are clearly well-described by a power law with a common slope for the various filters.  We fit a function only to the PSF size trend since it will be a more robust characteristic of the  survey, whereas ellipticity can have relatively wide spatial variations.  For a power law of the form
\begin{align}
\frac{\Delta R/ R}{\beta} = A_{\text{F}} 10^{B_{\text{F}} (m - m_{0})} \ (m_0 =20), 
\label{power_law}
\end{align}
the parameters $A_{\text{F}}$ and $B_{\text{F}}$ for each filter F are listed in Table 2. 

\subsection{Impacts of spatial variability of $\beta$}

\begin{figure}[!h]
\centering
\resizebox{\hsize}{!}{\includegraphics{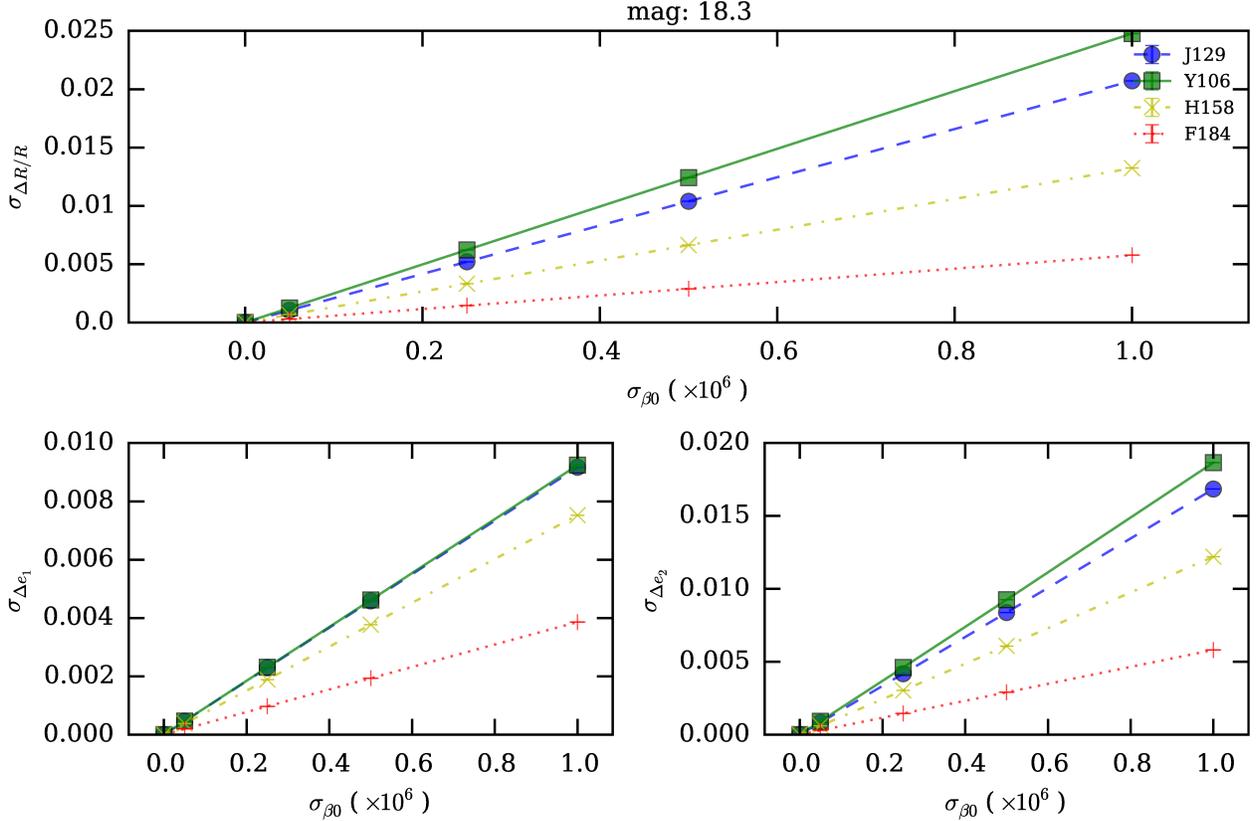}}
\caption{Impact of spatial variability in the $\beta_0$ parameter, quantified as the standard deviation over $M=100$ realizations (for a fixed centroid) in $\Delta e_1$,  $\Delta e_2$, and $\Delta R/R$  as a function of the standard deviation $ \sigma_{\beta_0}$. 
The source magnitude is fixed at 18.3.}
\label{f4}
\end{figure}

 We also studied the impact on $\Delta e$ and $\Delta R/R$ due to the dispersion in the $\beta$ parameter. Due to non-uniformities in fabrication, each pixel can have a different NL coefficient, and biases in the measurement of PSF properties could be introduced if a mean response curve is used to calibrate NL instead of a single curve per pixel. Fig. \ref{f4} shows the dispersion over $M=100$ realizations (for a fixed centroid) in the metrics $\Delta e_1$,  $\Delta e_2$, and  $\Delta R/R$  as a function of the standard deviation $ \sigma_{\beta_0}$, assuming that the coefficient $\beta$ is drawn from a distribution of the form $\mathcal{N} \sim (\beta_0, \sigma_{\beta_0})$. We have verified that the mean value of the metrics over all the realizations are consistent with the case where a fixed $\beta_0$ is considered, and therefore our stochastic beta model simply  introduces scatter around the shape measurement biases computed in Section 4.1.

\begin{figure}[!h]
\centering
\resizebox{\hsize}{!}{\includegraphics{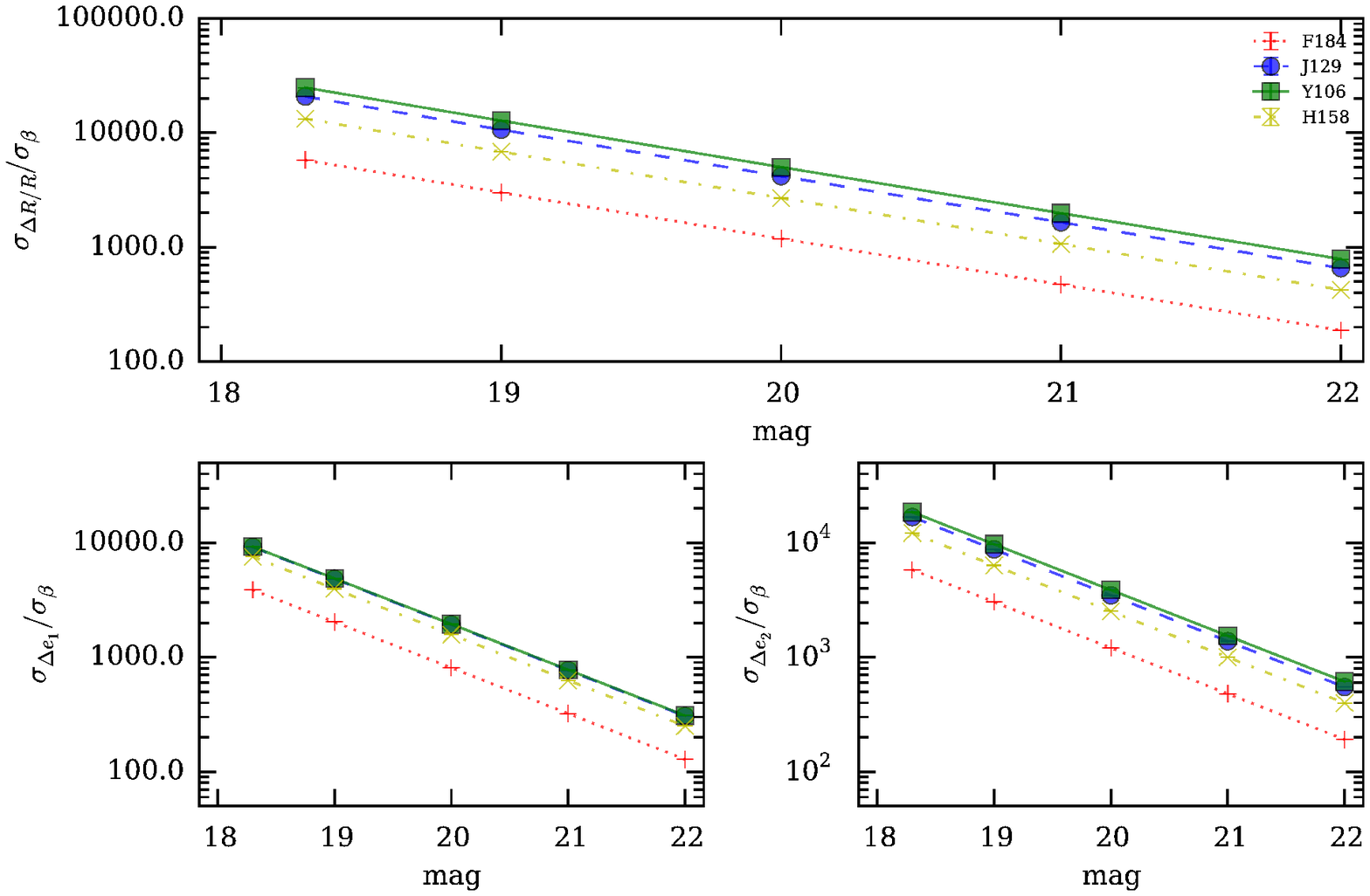}}
\caption{Standard deviation of the fractional error in PSF size and of the absolute error in PSF ellipticity components normalized by the error in the NL parameter $\sigma_\beta$, as a function of magnitude for the four HSL filters (J129, Y106, H158, and F184).}
\label{f5}
\end{figure}

We found that the results in Fig. \ref{f4} are insensitive to the nominal $\beta$ and therefore, in general, $\sigma_{\beta}$  can be treated as an error on $\beta$ estimation. The linear relationships in Fig. \ref{f4} also allows us to plot the slope of each curve for different magnitudes, as was done in Fig. \ref{f3}, and therefore analogous equations to Eq. \ref{calib} can be used to convert between the dispersion in errors on $\beta$ and the dispersions of errors in size and ellipticity measurements.  This computation may also be useful for deriving calibration requirements, although weak lensing analyses are much more sensitive to shape measurement biases than to small random errors.  The results are shown in Fig. \ref{f5}. We also fitted a model of the form 
\begin{align}
\frac{\sigma_{\Delta R/ R}} {\sigma_{\beta}} = A_{\text{F}}10^{B_{\text{F}} (m - m_{0})} \ (m_0 =20)
\end{align}
for the PSF size trends in Fig. \ref{f5}, as was done above for Fig. \ref{f3} (\emph{c.f.}, Eq. \ref{power_law}). The fit parameters are presented in Table 2 as well.  

\begin{table}
\centering
\begin{tabular}{|c|c|c|c|c|}
\hline
\multirow{2}{*}{Band} & \multicolumn{2}{c|}{size bias (Fig. \ref{f3})} & \multicolumn{2}{c|}{size scatter (Fig. \ref{f5})} \\
\cline{2-5}
& $A_{\text{F}}$ & $B_{\text{F}}$ & $\ A_{\text{F}}$ & $B_{\text{F}}$  \\
\hline
Y106 & 8322 & -0.4086 & 5057 &  -0.4045  \\
\hline
J129 & 6958 & -0.4089  & 4215 &  -0.4050 \\
\hline
H158 & 4668 & -0.4069 & 2711 &  -0.4038 \\
\hline
F184 & 2209 & -0.4035 & 1194 &  -0.4018 \\
\hline
\end{tabular}
\caption{Parameters resulting from fitting the curves in the upper panel of Figs. \ref{f3} and \ref{f5} to power-law functions of the form $\Delta R/ R/ \beta = A_{\text{F}}\ 10^{B_{\text{F}} (m - 20)}$ and $\sigma_{\Delta R/ R}/ \sigma_{\beta} = A_{\text{F}}\ 10^{B_{\text{F}} (m - 20)}$, respectively.}
\label{table2}
\end{table}

\section{Conclusion}
\label{Conclusion}
We have used the \wfa\ module in \gs\ to study the impact on PSF measurement for weak lensing science due to the nonlinearity in the conversion of charge to voltage in near-infrared hybrid CMOS detectors (such as those that will be used in the Wide Field Imager of the \wfa\ mission). The PSF profiles created by the {\tt{galsim.wfirst}} module posses several of the design characteristics of the expected PSF of the mission, such as optical aberrations and pixel scale. The module can also be used to assign the PSF profiles fluxes per pixel consistent with the expected brightness of the HLS. 

Voltage nonlinearity\textemdash as studied in this work\textemdash encompasses the linearity due to the shrinking of the depletion region at the p-n junction as charge accumulates, and the deviation from linearity originating in the multiplexer gain. It depends on the total integrated signal (fluence), and it is more dominant at high signals than other types of nonlinearity such as reciprocity failure, which dominates at lower signals. As such, NL will tend to depresses the flux in the core of the PSF relative to the wings in bright stars that are usually used for PSF estimation, introducing errors when deconvolving the PSF at the interpolated galaxy positions.

To model NL, we have used a single, one-parameter transfer function quadratic in the charge $Q$ (Eq. \ref{NL}). We have studied the consequences of NL in isolation by using the relationship in Eq.\ref{NL}, not considering other sensor effects, and neglecting sources of noise such as zodiacal background, thermal emission, and read noise (which would produce a negligible contribution). We have used the metrics $\Delta R/R$ and $\Delta e$ to assess the impact of NL on PSF size and ellipticity, which have to be controlled to $\sim O(10^{-4})$ or better, for different values of the parameter space at hand ($\beta$, PSF magnitude, and bandpass filters to be used in \wfa\ WL analyses).  We have also studied the effects of spatial variation of the $\beta$ coefficient along the pixel array, by assuming that it follows a Gaussian distribution. 

For the nominal value $\beta_0$ assumed in this work, we find that NL induces errors in PSF size and shape larger to what is tolerable by accurate WL measurements. For an example set of assumed requirements on PSF size and ellipticity ($10^{-4}$  and $4.7\times 10^{-5}$, respectively), we find that  $\beta$ should be calibrated to about $1\%$ to $2.4\%$ (H158 band). However, the results derived in this study (Eq. \ref{calib}, Fig. \ref{f3}, and Fig. \ref{f5}) can be used to derive requirements on NL for the \wfa\ detectors for a different set of tolerances on PSF properties. It is important to note that new measurements on the actual H4RG detector that will be used by the \wf\ imager will have to be performed in order to determine the mean value of $\beta$ and its dispersion.

Nonlinearity measurements are usually performed by looking at signal as a function of exposure time at a constant flux (and subtracting dark frames at the appropriate times), which is sensible to any gain dependence on fluence. These measurements are normally subject to other effects such as inaccuracies in the readout time, persistence, reciprocity failure, and time-dependent changes in the electronic offset due to self-heating effects in the multiplexer. Despite these challenges, the mean non-linearity signal can usually be characterized to a precision of 5\textendash10$\%$. Thus it is not obvious that a typical NL calibration program will be sufficient for WL with \wf\ without more careful study and error budgeting.
To ensure that voltage nonlinearity in the \wf\ H4RG detectors can be calibrated to the levels demanded by WL science, NL characterization will be crucial through the use of facilities such as the Detector Characterization Laboratory\footnote{https://detectors.gsfc.nasa.gov/DCL/} for rigorous calibration studies and the 
Precision Projector Laboratory (PPL, \citealt{seshadri13}, \citealt{shapiro13})\footnote{The PPL is a joint project between NASA Jet Propulsion Laboratory and Caltech which validates image sensor behavior using laboratory emulations of astronomical data.} for validating the impact of NL on shape measurement.\\

We thank Chris Hirata, Jeff Kruk, Dave Content, and the WFIRST detector requirements working group for useful discussions. AAP is supported by the Jet Propulsion Laboratory. CS and JR are being supported in part by the Jet Propulsion Laboratory. The research was carried out at the Jet Propulsion Laboratory, California Institute of Technology, under a contract with the National Aeronautics and Space Administration.

\copyright\ 2016. All rights reserved.

\acknowledgments
\bibliographystyle{plainnat}
\bibliography{ms}

\end{document}